\def\sqiglt{\hbox{\rlap{\lower.55ex \hbox{$\sim$}}\kern-.0em \raise.4ex \hbox{$<$}\,}}
\def\sqiggt{\hbox{\rlap{\lower.55ex \hbox{$\sim$}}\kern-.3em \raise.4ex \hbox{$>$}\,}}
\def\sqig{$\sim\,$} \def\etal{et\,al.} \def\msun{M$_{\scriptstyle\odot}$} 
  
 \def\pten#1{$\times10^{#1}$}
\def\deg{$^{\circ}$} \def\kmps{km\,s$^{-1}$} 
 \def\Hb{H$\beta$}
\def\Hg{H$\gamma$} \def\HeI{He\,{\sc i}} \def\HeII{He\,{\sc ii}} 
\def\HeIIl{He\,{\sc ii}\,$\lambda$4686} 
\def\minone{$^{-1}$} \def\kev{\,ke\kern-.1em V}
\def\beat{$\omega$\,--\,$\Omega$}\def\voph{V2400~Oph}
\def\appro{$\approx\,$}\def\rmag{$R_{\rm mag}$}\def\rwd{$R_{\rm wd}$}
\def\rcirc{$R_{\rm circ}$}\def\rmin{$R_{\rm min}$}\def\rinner{$R_{\rm inner}$}
\documentstyle{mn}
\title[The intermediate polar V2400~Oph]
{The accretion flow in the discless intermediate polar V2400~Ophiuchi}
\author[C. Hellier and A.\,P.\,Beardmore]
{Coel Hellier and A.\,P.\,Beardmore\\
Astrophysics Group, School of Chemistry and Physics, Keele University, 
Keele, Staffordshire, ST5 5BG}
\date{Accepted ???. Received ???}
\begin{document}
\maketitle
\begin{abstract}
{\sl RXTE\/} observations confirm that the X-ray lightcurve of \voph\
is pulsed at the beat cycle, as expected in a discless intermediate
polar. There are no X-ray modulations at the orbital or spin cycles,
but optical line profiles vary with all three cycles.  We construct a model
for line-profile variations in a discless accretor, based on the idea
that the accretion stream flips from one magnetic pole to the other,
and show that this accounts for the observed behaviour over the spin
and beat cycles. The minimal variability over the orbital cycle
implies that 1) \voph\ is at an inclination of only \appro 10\deg, and
2) much of the accretion flow is not in a coherent stream, but is
circling the white dwarf, possibly as a ring of denser, diamagnetic
blobs. We discuss the light this sheds on disc formation in
intermediate polars.
\end{abstract}
 
\section{Introduction}
The magnetic cataclysmic variables can be divided into two classes. In
polars (or AM~Her stars) the magnetic field of the white dwarf is
strong enough to lock its rotation to the binary orbit, and accretion
proceeds via a stream which is deflected by the field onto a magnetic
pole. In contrast, in intermediate polars (IPs or DQ~Her stars) a
lower-field white dwarf spins more rapidly than the orbit, and the
stream feeds into an accretion disc, which then feeds field lines from
its inner edge (see Warner 1995 for a comprehensive review).

The distinction is blurred, however, by the existence of several 
asynchronous polars, in which the spin periods differ from the orbit
by \sqig 1 per cent (e.g.\ Schwope \etal\ 1997). There has also been
a long debate on the existence of discless IPs (e.g.\ Hameury,
King \&\ Lasota 1986; 
King \& Lasota 1991; Hellier 1991; Wynn \&\ King 1992).  In such 
systems, the stream would be expected to flip first to the upper pole,
then to the lower pole, as the magnetic dipole rotates (e.g.\ 
Hellier 1991; Wynn \&\ King 1992).
Such pole flipping would occur at the frequency with which the relative geometry
changes, namely the beat frequency \beat, where $\Omega$ and $\omega$
are the orbital and spin frequencies respectively.

Buckley \etal\ (1995; 1997) reported the first secure evidence for a
discless IP, with the discovery of the {\it Rosat\/}
source \voph\ (RX\,J1712.6--2414). They found that polarised light
from the system varies at 927 s, which is interpreted as the spin
period of the magnetic dipole and thus of the white dwarf.  The
X-rays, though, are pulsed at 1003 s, which is the beat period between
the 3.42-h orbital and 927-s spin cycles; thus \voph\ shows the signature 
of pole-flipping accretion, as expected in a discless IP.

\voph\ is a prime opportunity to study the interaction of an
accretion stream with a magnetic field in a situation where the flow
is continually changing as the dipole rotates. This contrasts with
most magnetic cataclysmic variables where the flow is expected to
settle (at least temporarily) into a quasi-equilibrium.  (In principle
the asynchronous polars offer the same opportunity, but it is much
harder to obtain good coverage of their \sqig 50-d beat cycles.)  In
this paper we present new X-ray observations of \voph, along with
optical spectroscopy aimed at tracing the accretion flow as it 
connects to the field lines.

\begin{figure*}\vspace*{5.5cm}     % Fig 1 
\caption{The 2--15-\kev\ X-ray lightcurve of \voph\ as recorded by 
{\sl RXTE}.}
\includegraphics{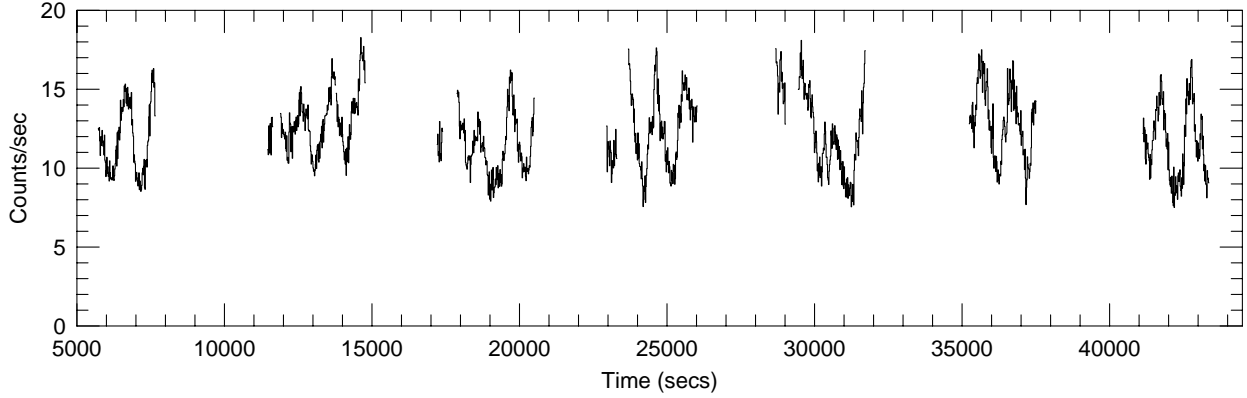} 
\end{figure*}

\begin{figure*}\vspace*{4.8cm}     % Fig 2
\caption{The Fourier transform of the {\sl RXTE\/} observation, part of
which is shown in Fig.~1. The orbital ($\Omega$) and spin ($\omega$) 
frequencies are marked. The dominant modulation, with an alias structure
caused by the spacecraft orbit, is at the beat frequency, \beat.}
\includegraphics{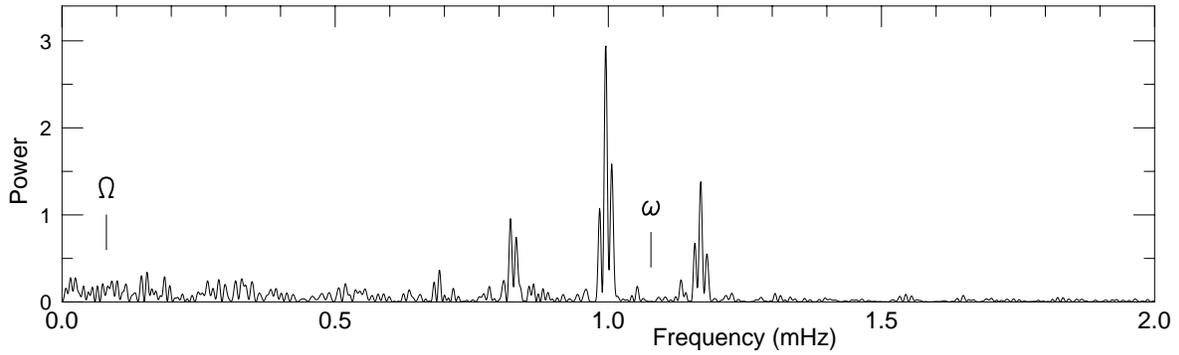} 
\end{figure*}

\voph\ is the most polarised IP, leading to a magnetic field 
estimate of 9--27 MG (Buckley \etal\ 1995; V\"ath 1997), the highest
for any IP and overlapping with the range for low-field polars. The
high field may explain why a disc has not formed in this system,
whereas they do in IPs at \sqig 1 MG.  Buckley \etal\ (1995) also
found that the circular polarisation is always of the same sign, and
so concluded that we only ever see one magnetic pole.
Given, also, a lack of radial velocity motion at the orbital
period, they proposed that \voph\ is at a low inclination, and that we
only see the pole nearest us (the `upper' pole). From the fact that
there were no detectable variations in linear polarisation, Buckley
\etal\ (1995) suggested that this pole was never viewed side on, and
so proposed that both the inclination of the binary, and the angle
between the spin and magnetic axes of the white dwarf, $\delta$, are low.

\begin{figure}\vspace*{7.0cm}     % Fig 3
\caption{The X-ray lightcurve of \voph\ folded on the 1003-s beat period.
A typical error is shown.}
\includegraphics{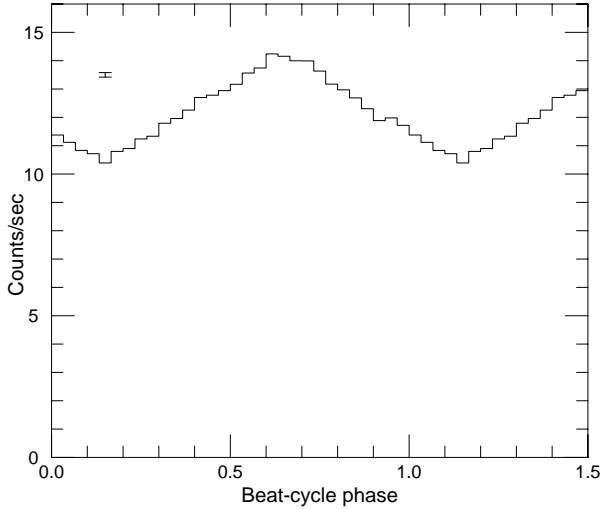} 
\end{figure}

\begin{figure}\vspace*{4.2cm}     % Fig 4 
\includegraphics{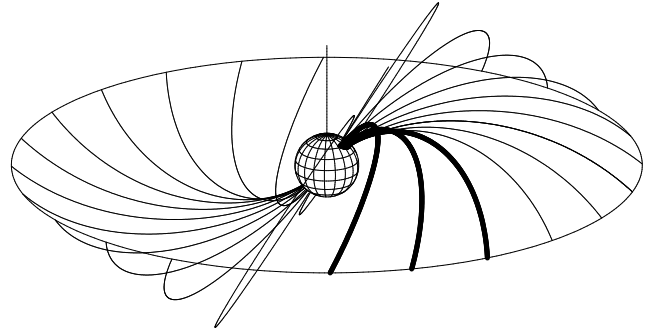} 
\caption{An illustration of the discless accretion geometry in \voph.
In our model (Section 4) the stream attaches to a 20\deg\ swathe of
field lines (shown in bold). The change in geometry over the beat cycle can
be visualised by imagining the feeding point moving round the ring.}
\end{figure}

\section{An {\sl RXTE\/} observation}
We report, first, on an {\sl RXTE\/} X-ray observation obtained over the
interval 2000 June 24--25. This resulted in 39 ks of data from the 
PCA instrument, with either 3 or 4 of the 5 PCU modules in
operation at any time. A section of the resulting lightcurve is shown
in Fig.~1, while the Fourier transform of the entire observation is
shown in Fig.~2.  Both lightcurve and transform are dominated by a
pulsation at the 1003-s beat period; this has a maximum amplitude of
\appro 50 per cent [(peak--trough)/peak] and a mean amplitude of 25
per cent, as shown by the folded pulse profile in Fig.~3.  There is no
sign of variations at the orbital period ($<$\,7 per cent) or at the spin
period ($<$\,2 per cent).  A preliminary look at five other {\sl RXTE\/}
observations, of similar length and spaced at intervals of a few months,
finds the same behaviour:  a strong beat-cycle pulse but no modulation
at the orbital or spin cycles.

\begin{figure*}\vspace*{5.5cm}     % Fig 5
\caption{The summed spectrum of \voph.}
\includegraphics{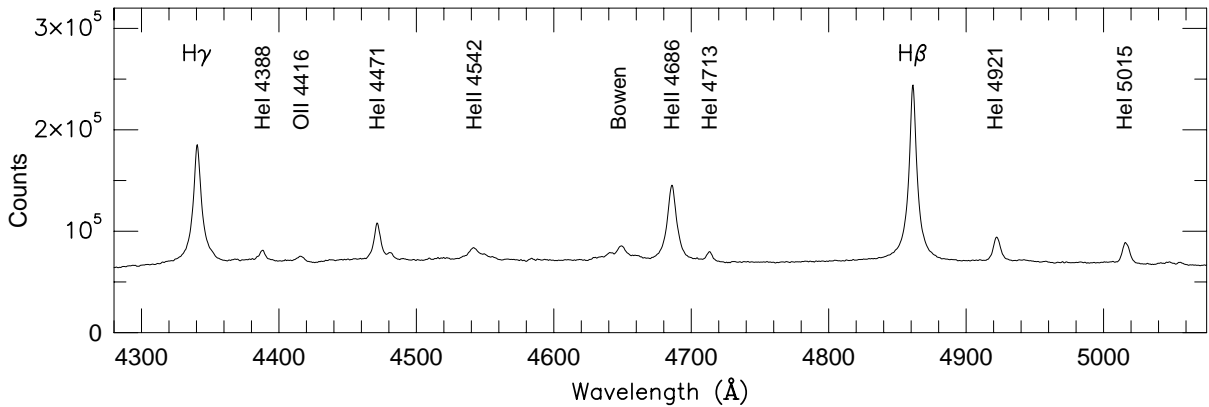} 
\end{figure*}

The absence of an orbital modulation in the X-ray lightcurve is
unexpected in a discless IP. If the stream flows to the magnetosphere,
latches onto the nearest field line, and follows it onto the white
dwarf (see Fig.~4), then the accretion sites will always be on the
hemisphere facing the secondary, and thus their visibility should vary
over the orbit, as in an AM~Her star. The lack of an orbital
modulation can, though, be explained if we concur with the suggestion that
\voph\ is at so low an inclination that the upper pole is always in
view and the lower pole never in view.

The next puzzle arises from the fact that the beat-cycle modulation is
not total. If all the accreting material participates in the pole
flipping, then, given the low inclination, the flow would flip to
the far, hidden hemisphere of the white dwarf for roughly half the
beat cycle, leading to 100 per cent modulation. If some of the flow
does not flip, but manages to feed both poles continually, then this
could explain the non-total modulation. But why, then, does this
not result in a spin-cycle pulsation, as seen in all the IPs where a
disc ensures continual feeding of both poles?  One answer could be to
invoke a field aligned exactly with the spin axis, $\delta = 0$,
but this would also
remove the beat pulse, since nothing would change with beat phase.  It
also conflicts with the fact that we see spin-cycle variations of the emission
lines (see Section 7).

A second answer might be to invoke large, extended accretion
footprints, coupled with a large $\delta$ so that these sites would be 
near the white-dwarf equator. In a face-on system, any accretion sites
extending across the equator would always be partly in view, and, with
both poles partially visible, any pole-flipping modulation would not
be total.  Arguing against this, however, is the fact that the
visibility of the poles would then be exquisitely sensitive to orbital
phase, and we do not see any orbital modulation in the X-ray
flux. Also, the fact that we see variable circular polarisation but
not variable linear polarisation argues that we do not see the poles
sideways on (Buckley \etal\ 1995).

A final possibility could invoke a field geometry that is more complex
than a dipole, in which both poles were always on the near, visible
hemisphere. However, this is contradicted by the fact that the
circular polarisation always has the same sign, arguing that we only
see one pole (Buckley \etal\ 1995). Thus, the absence of an X-ray spin
pulse remains a puzzle, which we defer to Section 8, where we
suggest a possible explanation.

\section{Spectroscopic observations}
We observed \voph\ with the 3.9-m AAT and the RGO spectrograph plus a
TEK CCD. A 1200 lines mm\minone\ grating gave a resolution of 1.4\AA,
covering the range \Hg\ to \Hb. Observing for 5.0 h, 4.5 h \&\ 4.5 hr
on the three consecutive nights 1996 May 10--12 we obtained 900
integrations of 50 s each, thus covering \appro 4 orbital cycles and
\appro 50 spin cycle of the star.  The summed spectrum, containing
\HeI\ and \HeII\ lines in addition to the Balmer lines, is shown in
Fig.~5.

\begin{figure*}\vspace*{13cm}     % Fig 6
\caption{The Fourier transforms of the equivalent widths and V/R ratios
of the \Hb\ and \HeIIl\ lines.}
% The orbital ($\Omega$) and spin ($\omega$) frequencies are marked.}
\includegraphics{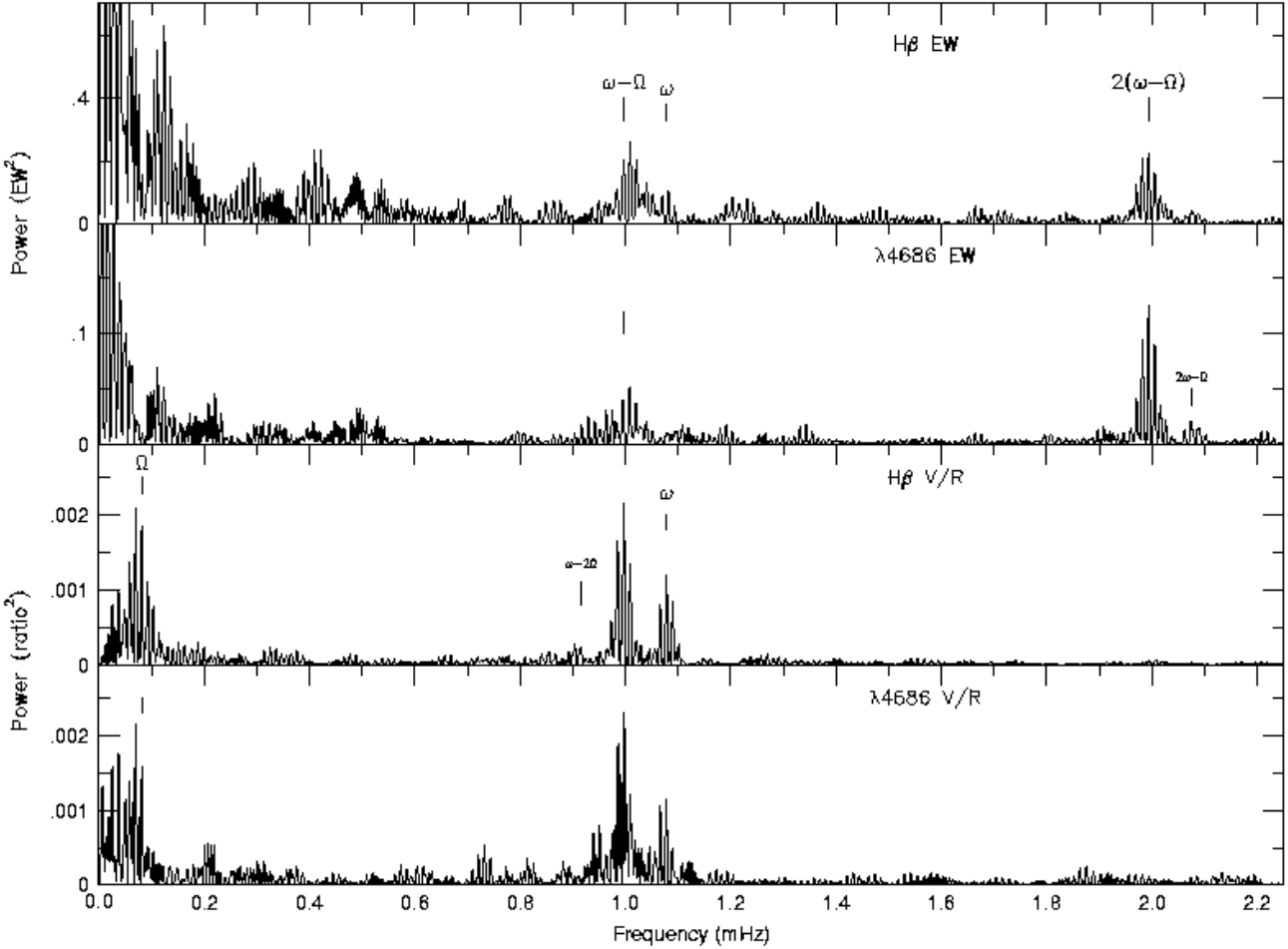} 
\end{figure*}

As a first look at the data we computed the equivalent widths and the
V/R ratios for the lines in each spectrum (V/R being the ratio of the
equivalent widths on either side of the rest wavelength).  The Fourier
transforms of these quantities for \Hb\ and \HeIIl\ are shown in
Fig.~6.

All four Fourier transforms show significant power at the beat (\beat)
frequency, which is always stronger than the power at the spin
frequency ($\omega$). A harmonic of the beat frequency, 2(\beat), is
also prominent in the equivalent widths. Other sidebands detected
include $\omega$\,--\,2$\Omega$ in the \Hb\ V/R ratios and
2$\omega$\,--\,$\Omega$ in the \HeIIl\ equivalent widths. There is
significant power at the orbital frequency in the V/R ratios (a
one-day alias being the highest peak) but not in the equivalent
widths.

To help us interpret the line profiles of \voph\ we have computed 
simulations of the profiles expected in a discless IP. We first 
describe this model, and then present a comparison of simulated and
observed profiles.

\section{Model line profiles}
The model we use to simulate the line profiles of \voph\ is
illustrated in Fig.~4. We assume a dipole field centered at the
white-dwarf centre --- observations of AM Her stars often indicate the
need for a more complex field (e.g.\ Wickramasinghe \&\ Ferrario 2000) 
but we know less about the fields in IPs, so adopt the simplest
approach. We assume that at some radius, \rmag, the stream attaches
to the nearest field line and follows this line, taking the
gravitationally downhill direction, onto the white dwarf. (Strictly,
the field strength, and thus \rmag, would be a function of the dipole
orientation, but how the material feeds onto the field lines, whether
it passes through a shock, and how much the rate of feeding depends
on dipole orientation are all uncertain, so again we adopt the 
simplest approach.) 

\begin{figure}\vspace*{22cm}     % Fig 7\vspace*{12cm}     % Fig 7
\caption{Simulated line profiles in a time sequence covering two
orbital cycles. The four panels show profiles for different distances
from the white dwarf. Darker colouring signifies stronger emission in all
figures.}
\includegraphics{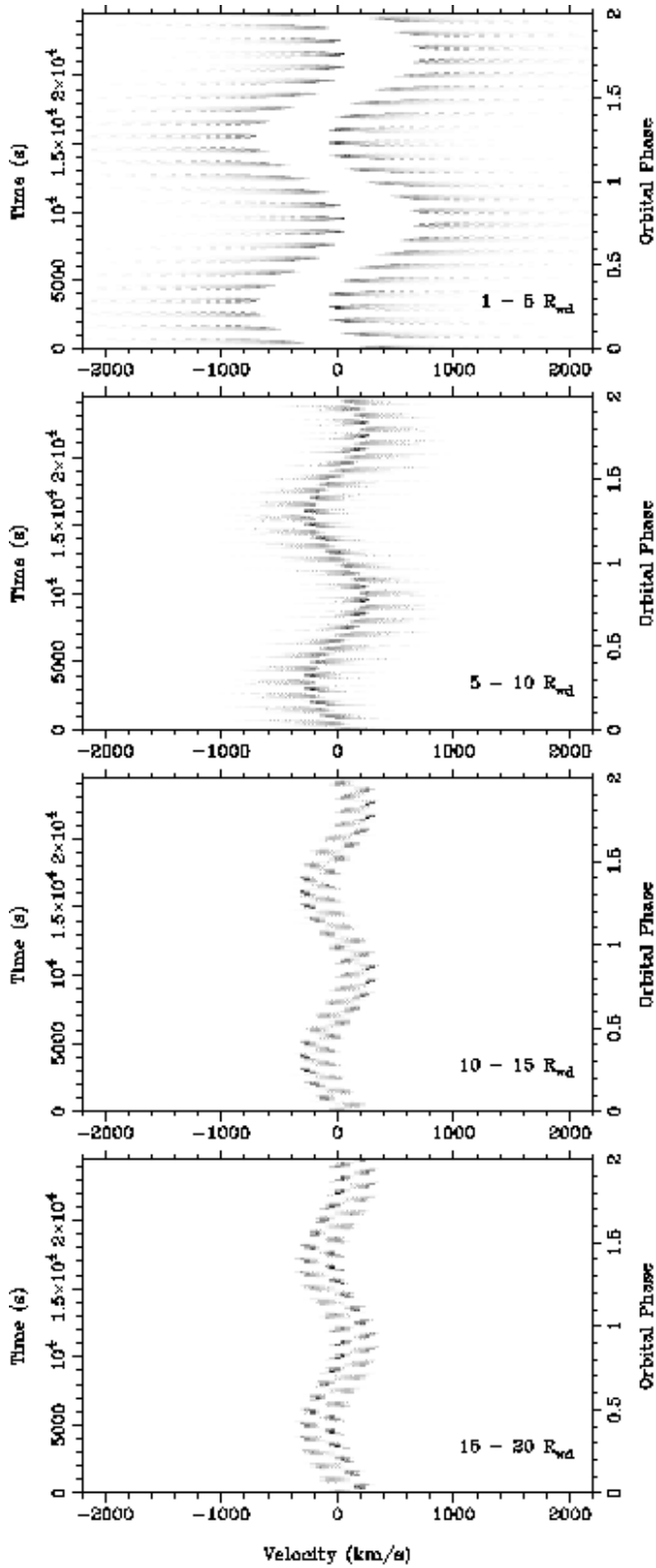} 
\end{figure}

The radius \rmag, the angle $\delta$, and the inclination of the binary 
are all free parameters.  We assume, arbitrarily, that the stream
feeds a 20\deg\ swathe of azimuth.  We then calculate the free-fall
velocities along the field lines and project them onto the line of
sight to produce line profiles. We assume that the emitting regions
are optically thick, and so scale the line intensity by projected
area.  Also, for maximum simplicity, we assume that the field lines
are not distorted by their interaction with the flow, and that the
material starts infalling with near-zero velocity at \rmag, as though
it passes through a shock.  We also assume that the infall time is short 
compared to the spin cycle, and thus do not allow for the fact that the 
stream will have moved on by the time the material accretes. 
Lastly, we rotate the dipole and stream
location to mimic the spin and orbital cycles. We do not include any
emission from the stream itself, further out than \rmag, which would
produce a lower-velocity S-wave on the orbital cycle.  Note that
the above model is very similar to that developed by Ferrario \&\
Wickramasinghe (1999), the main difference being that they aimed for
model Fourier transforms whereas we aim to compare line profiles.

Fig.~7 shows a sample set of simulated line profiles. We adopted spin 
and orbital cycles of 927 s and 3.42 h, a white dwarf mass of 0.80 \msun, 
and a secondary mass of 0.37 \msun. The inclination was 10\deg, the 
dipole-offset, $\delta$, was 60\deg\ and the radius \rmag\ was 20 \rwd. 
The four panels are for emission at different distances from the white 
dwarf.  

A velocity variation at the orbital cycle is obvious in all panels of
Fig.~7. Note that this is
primarily the infall velocity of the stream, as in an AM Her star, and
is not the orbital velocity of the white dwarf, which is much smaller.
The spin and beat cycles are seen as the much faster variation in the
simulated profiles. 

\begin{figure}\vspace*{22.2cm}   %\vspace*{12cm}     % Fig 8  % Fig 8
\includegraphics{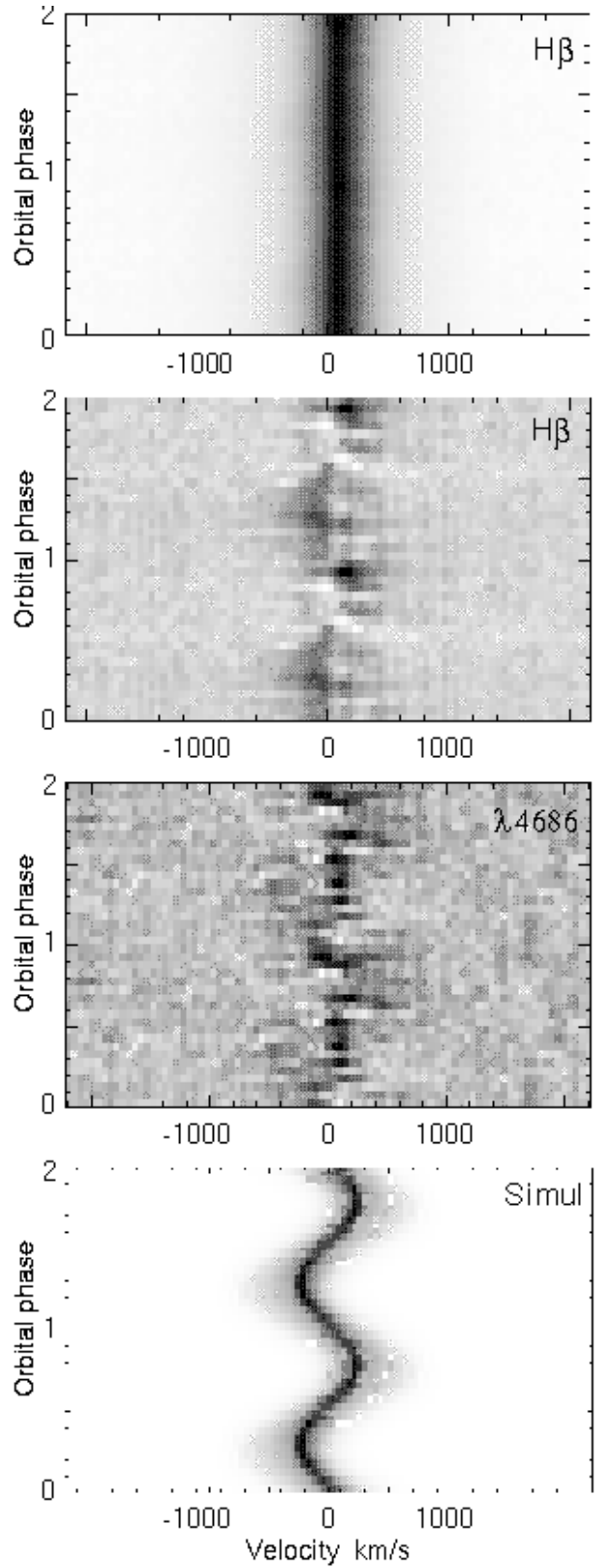} 
\caption{Line profiles folded on the orbital period. The top panel shows
the \Hb\ line; below it are the \Hb\ and \HeIIl\ lines after subtraction 
of the phase-invariant profile. At bottom are the simulated profiles 
for 5--10 \rwd.}
\end{figure}

\section{Line profiles over the orbital cycle}
The Fourier transforms (Fig.~6) revealed a variation of the line
profiles over the orbital cycle, but when the observed profiles are
folded on the orbital cycle it is hardly visible (Fig.~8).  To enhance
the variation we have subtracted the phase-invariant profile (for each
velocity we found the phase bin with the lowest value and subtracted that
from the data). In the portion of the line remaining, only 5 per cent of the
original, the slight variation is discernable. Since the orbital
modulation is so small, it is clear that \voph\ must be at a
very low inclination (as already suggested by the X-ray data).
But even so, the lack of variation is clearly discrepant with the model,
which is calculated for $i = 10$\deg. 

\begin{figure}\vspace*{22cm}     % Fig 9
\includegraphics{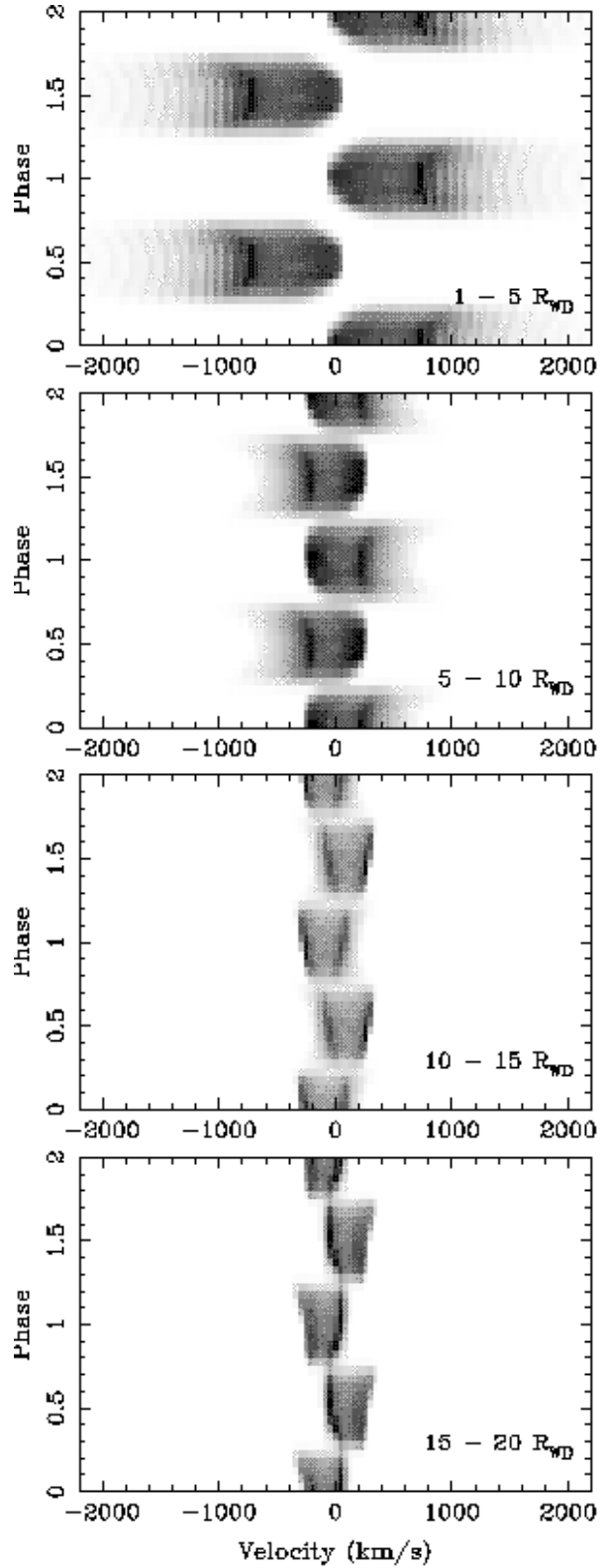} 
\caption{The simulated line profiles folded on the beat period. 
The four panels correspond to different distances from the white dwarf.}
\end{figure}

By fitting Gaussians to the phase-resolved \Hb\ profiles we measure a
radial-velocity variation of only 5.8 \kmps, which agrees with Buckley
\etal\ (1995) who report a limit of $<$\,10 \kmps.
If this is the motion of the
white dwarf it implies an inclination of only 3\deg\ (assuming stellar
masses of 0.7 and 0.3 \msun). Even this could be too high, 
since, as shown by the simulations, we expect the orbital motion to be
dwarfed by an infall velocity of anything up to the white-dwarf escape
velocity. The observations and simulations can be
reconciled only if 1) the system is at an inclination of $<$ 1\deg\ 
(unlikely, since the {\it a priori\/} probability is only 1.5\pten{-4}), 
or 2) much of the
emission comes from material that has circularised about the white
dwarf and dilutes the emission from the infalling stream.  Thus the
variations in the emission lines might be tracing only a part
(possibly a small one) of the flow.

\section{Line profiles over the beat cycle}
Fig.~9 shows the simulated profiles folded on the beat period, again for
four distances from the white dwarf. The changes in velocity as the 
stream changes direction to flow to a different pole are obvious. 
Note that, at such a low inclination, the dominant component of the
velocity is the motion out of the plane following the arch of the
field lines.  At phase zero the stream is flowing to the upper pole,
and the emission 10--20 \rwd\ out is blueshifted, since it is rising 
out of the plane towards us; the emission 1--10 \rwd\ out is falling
back to the plane and so is redshifted. 

\begin{figure}\vspace*{17cm}     % Fig 10
\includegraphics{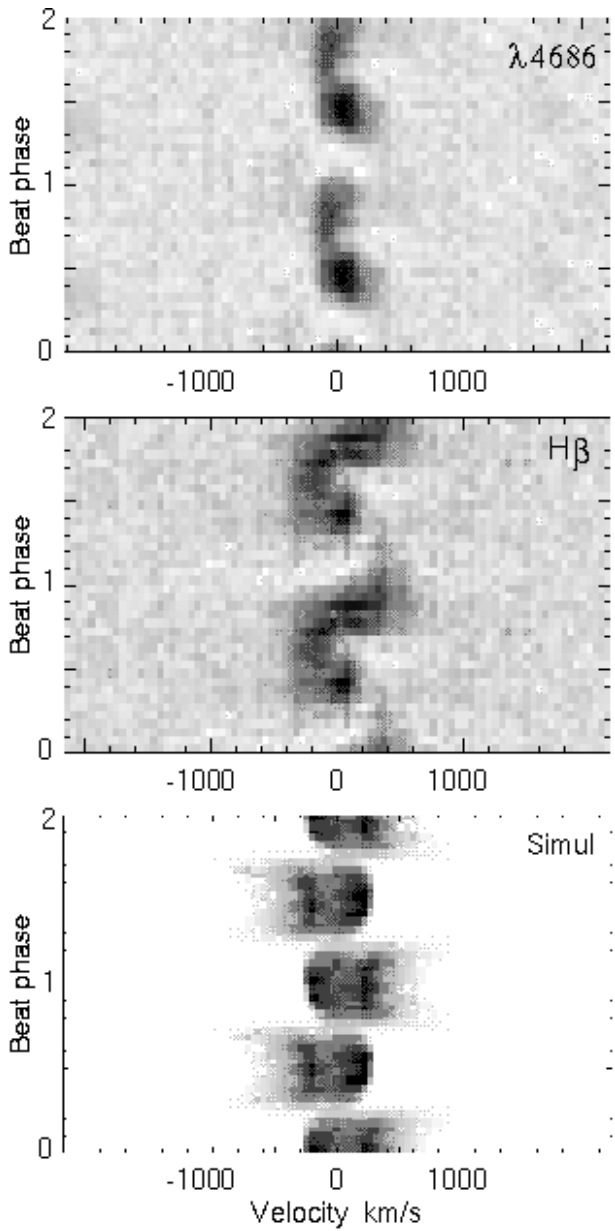} 
\caption{The observed and simulated line profiles folded on the beat period. 
The data are shown after subtraction of the phase-invariant profile.
The relative phasing of data to simulation is arbitrary.}
\end{figure}

In Fig.~10 we show the observed profiles of \Hb\ and \HeIIl\ folded on 
the beat cycle.
Again, we have first subtracted the phase-invariant profile (leaving
only 5 and 10 per cent of the two lines, respectively), since the
variation is barely discernable in the raw fold. Accompanying the data
are the simulated profiles from 5--10 \rwd\ out, which give a rough 
match. The similarity of the \Hb\ line to the simulation, with both
showing changes from red shift to blue shift,  suggests that
we are seeing the change in velocity as the stream flips from pole
to pole. In the observations, however, the flip
is less abrupt than in the model, perhaps indicating that the stream
feeds field lines over a larger range of azimuth than the 20\deg\
adopted in the model.
The \HeIIl\ line does not match so well, and shows little variation
in velocity. Possibly this line comes mostly from near the apex 
of the arch, where motion out of the plane is minimal.

Our model predicts that the line intensity is modulated at twice
the beat frequency (as did the model of Ferrario \&\ Wickramasinghe
1999), and this agrees with the presence of a 2(\beat) modulation in
the equivalent widths (Fig.~6).  The weaker modulation seen in the
equivalent widths at \beat\ is not predicted by our model, which
has symmetric upper and lower poles, but would result
from any asymmetry between the poles. Another factor is that X-ray
illumination of the stream will vary with beat phase, and this would
also modulate the equivalent widths at the beat period.

\begin{figure}\vspace*{22cm}     % Fig 11
\includegraphics{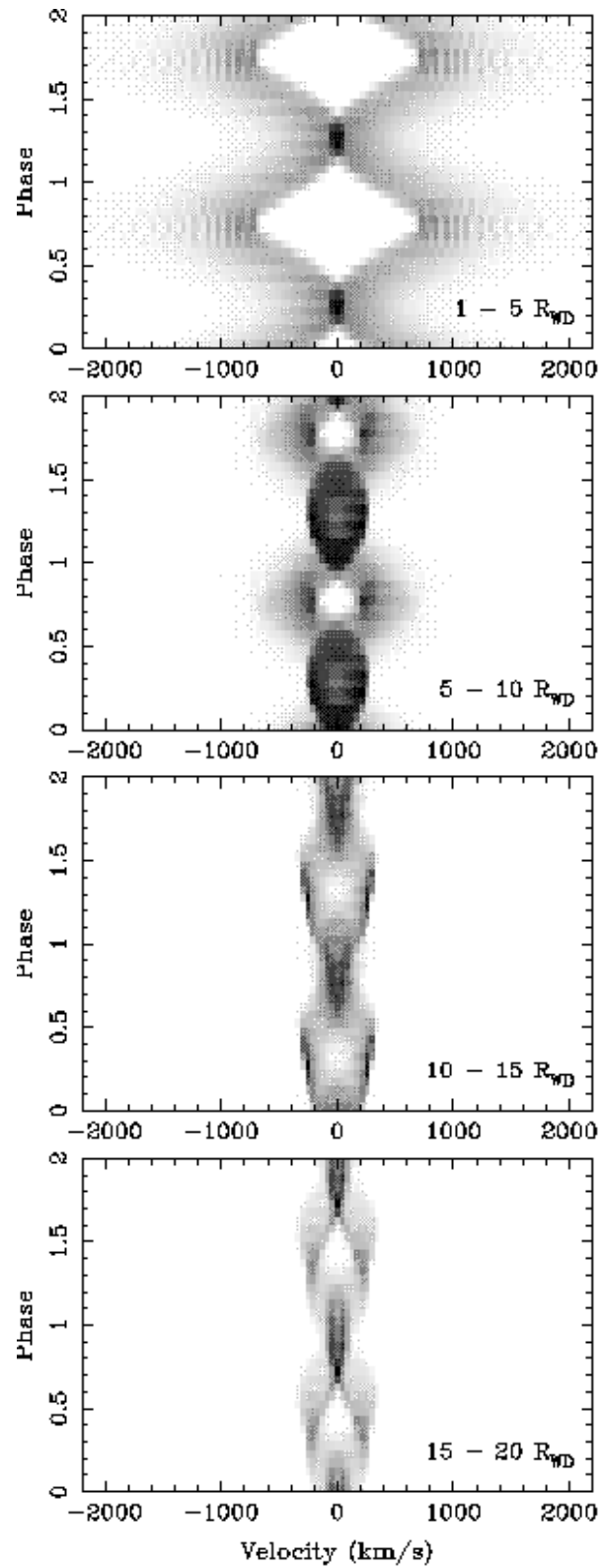} 
\caption{The simulated line profiles folded on the spin period. 
The four panels correspond to different distances from the white dwarf.}
\end{figure}

\section{Line profiles over the spin cycle}
Fig.~11 shows the simulated profiles folded on the spin cycle. 
The symmetric shape is the result of the two opposite poles, but
note that at any one time only one is accreting. Thus a `snaphot'
of the simulation would show an S-wave, snaking from red to blue
over the spin cycle; after a pole-flip the inverse S-wave would be
seen, and such S-waves accumulate to produce the symmetric pattern
when a long sequence is folded.

Again, the dominant velocity is the motion out of the plane, so that
the profiles 10--20 \rwd\ out (where material rises out of the plane)
tend to show the reverse of the profiles from 1--10 \rwd\ (where material
falls back). If there were no effect of inclination the profiles 
would repeat exactly twice per spin cycle, but the tilt to an 
inclination of 10\deg\ is enough to break the symmetry.

Fig.~12 shows the spin-folded \Hb\ and \HeIIl\ profiles, again 
accompanied with the simulated profiles from 5--10 \rwd, which give
a good match to the \HeIIl\ line.  In the simulation
the upper pole is at the ``3 o'clock'' position at phase zero and rotates
counterclockwise.
At phases 0.2--0.3 we are seeing the upper pole furthest from us and the
lower pole towards us.  The material at 5--10 \rwd\ is beginning to fall
back to the plane. The 10\deg\ inclination tilts this motion towards the 
plane of the sky, reducing the projected velocities, and the profile
is thus narrowest. Half a cycle later (upper pole towards us), the 
inclination tilts the infall motion towards the line-of-sight, and the 
observed velocities are higher. 

The profiles from \Hb\ are noisier than those from \HeIIl: after
subtraction of the phase-invariant profile only 3 per cent of the 
original \Hb\ line remained. However, the profile can be interpreted
as being similar to that of \HeIIl, with the exception that the 
blue wing seen at phases 0.4--0.8 is missing.   

\begin{figure}\vspace*{17cm}     % Fig 12
\includegraphics{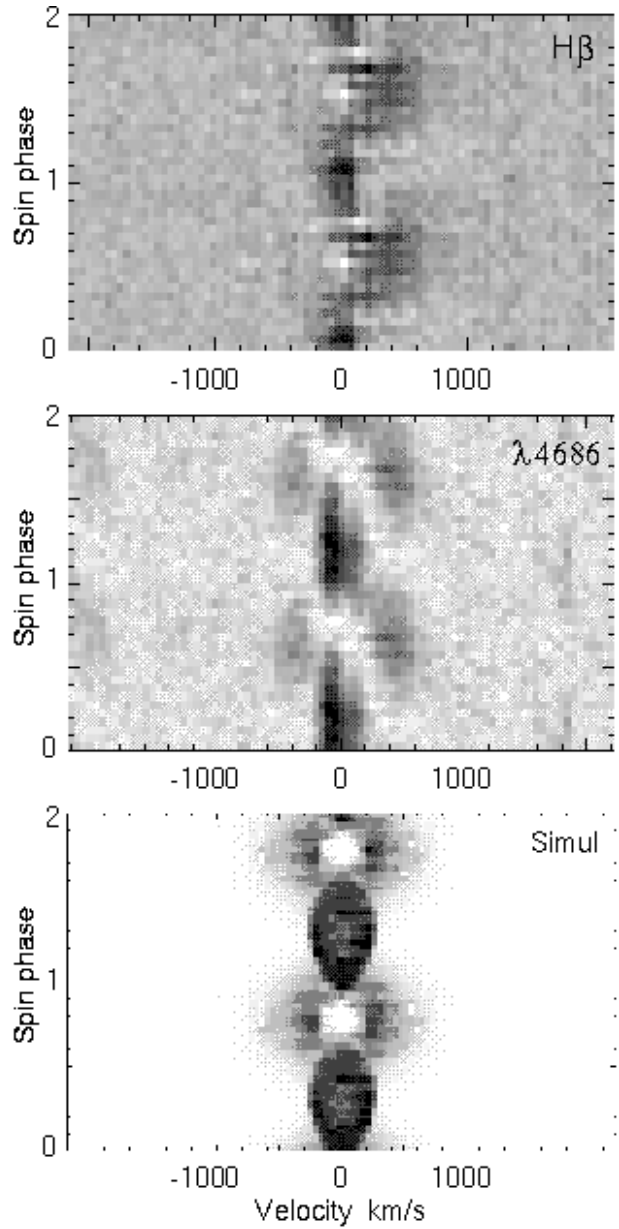} 
\caption{The observed and simulated line profiles folded on the spin period. 
The data are shown after subtraction of the phase-invariant profile.
The relative phasing of data to simulation is arbitrary.}
\end{figure}

\subsection{Varying the model parameters}
In the previous sections we have compared observed line profiles with
models calculated for emission 5--10 \rwd\ from the white dwarf.  We
should emphasise that we do not regard the models as fits, since the
model contains too many assumptions and adjustable parameters to lead
to a unique solution. Instead, we describe here the result of varying
the parameters. Since the simulation calculates infall
velocities, changing \rmag\ or the white-dwarf mass simply scales the
velocities without altering the shape of the model profiles. Similarly, 
adding an injection velocity at \rmag\ would increase the velocities
at all points. Increasing the 20\deg\ range of azimuth over which accretion
occurs would smear out the profiles, but retain the overall shape.

However, the shape of the spin-resolved profiles is affected by the
dipole offset, combined with the inclination.  At zero inclination, of
course, our view would not change with spin phase and there would be
no spin-cycle modulation. This rules out the extreme inclinations of
$<$\,3\deg\ discussed in Section 5 --- only by adopting an
inclination of at least 10\deg\ do we reproduce the observed
spin-resolved profiles, in which the narrow profile at phase \appro 0.2
changes to a broader, split profile at phase \appro 0.7.  Even with an
inclination of 10\deg, the profile is only reproduced with 
$\delta$\,$>$\,30\deg: any lower and the change in viewing angle over
the spin cycle is insufficient to reproduce the observed profiles.  Of
course, we can also reproduce the spin-resolved profiles using a
higher inclination, but that only increases the problem of the lack of
orbital modulation.

\section{Discussion}
We have shown that the emission lines in \voph\ vary on both the spin
and the beat cycles, and that the variations are compatible with a
model in which the accretion flow flips from one magnetic pole to the
other over the beat cycle --- the hallmark of discless accretion in an
asynchronous magnetic cataclysmic variable.  We have found that a
simple model of this behaviour, in which a stream falls to the 
magnetosphere and then feeds the nearest field line,
reproduces the observed line-profile changes over the spin and beat
cycles. However, it should be noted that the line-profile variations
involve only \appro 5 per cent of the line emission, and that the
bulk of the emission shows no detectable variation.

As a result of this, the variations over the orbital cycle are 
harder to interpret. We adopt an inclination of 10\deg, since
this is the minimum necessary to reproduce the line-profile variations
over the spin cycle (Section 7). However, if all the emission came
from a collimated, infalling stream, this inclination would lead to  
an orbital modulation of hundreds of \kmps, whereas the observed
value is only 5.8 \kmps\ (Section 5). This implies that the bulk
of the emitting material is not in the infalling stream, but is
instead symmetric about the white dwarf. 

This interpretation is supported by X-ray data, in which the
beat-cycle modulation is only 25 per cent deep. Given that the system
is at so low an inclination that we only ever see the upper pole,
this suggests that only a minority of the accreting material flips from
pole to pole, and that there is a continual flow to both poles irrespective
of beat-cycle phase.  

What is the nature of this flow?  One possibility might be a conventional
accretion disc. This could feed field lines from its inner edge, forming
`accretion curtains' of material in the usual manner for an IP. However,
why, then, do we see no X-ray spin pulse? Given that the inclination and
dipole offset are sufficient to produce obvious spin-cycle variations
in the emission lines, we would expect the aspect of the accretion sites
to vary with spin phase, and thus absorption and scattering of X-rays
in the accretion curtain would produce a spin pulse (this is best 
documented in AO~Psc; Hellier, Cropper \&\ Mason 1991; Hellier \etal\
1996; see also the model of Kim \&\ Beuermann 1995).
Another issue is how the stream coexists with the disc, since, having
different velocities, they couldn't both occupy the orbital plane. 
One plausible answer is that the stream could overflow the disc (see 
Armitage \&\ Livio 1998 for theoretical simulations of this). This idea
has been invoked to explain the IPs showing both spin and beat pulses in the
X-ray lightcurves (e.g.\ Hellier 1991).
However, again, a spin-cycle pulsation is nearly always seen, and, 
taking the example of FO~Aqr, the beat-cycle modulation is always 
weaker or even absent (Norton \etal\ 1992; Hellier 1993b; 
Beardmore \etal\ 1998).

In view of the difficulties in invoking a disc, we turn instead to the
idea of magnetically threaded accretion flows, developed by King (1993)
and Wynn \&\ King (1995).
This model treats the flow as diamagnetic blobs, which are diverted from
a ballistic trajectory by the magnetic drag produced by their crossing
field lines.  The resulting trajectories depend on blob size and density.
Simulations by King \&\ Wynn (1999) suggest that it is possible
for less-dense matter to be easily threaded and controlled by field lines
(which would then act as in our simulations),  while denser
blobs cross field lines and circulate around the white dwarf. These blobs
would lose knowledge of orbital phase, and eventually be destroyed by 
magnetohydrodynamic instabilities, feeding their material onto the white 
dwarf's poles. 

Back to the same question, why doesn't this result in an
X-ray spin pulse? We speculate that this process is much less orderly
than in a disc-fed accretor.  A disc might supply a steady flow over
a relatively small disc--magnetosphere transition region, and so feed
accretion sites that are small and localised. For example they cover 
only $<$\,0.002
of the white-dwarf surface in XY~Ari (Hellier 1997b). This would lead 
to steady, well defined accretion curtains, able to absorb or scatter 
the X-ray flux and so modulate it over the spin cycle. In the 
diamagnetic-blob model, the accretion sites could be widely strewn 
and transient; they may not form coherent accretion curtains and so the 
spin-cycle modulation could be reduced to below the detection threshold.

\subsection{Relation to other systems}
We have detected variations of the line profiles over the spin
frequency in \voph, whereas hitherto the spin frequency in this star 
has been seen only in circular polarimetry. This is reassuring, for
the following reason.  Most IPs show little or no polarised light, and
in such systems an X-ray pulse is typically interpreted as
the spin period.  But if it was actually the beat period, how could we
know?  The fact that both spin and beat frequencies show up in the
line profiles of a discless accretor like \voph\ means that we can
distinguish between the spin and beat frequencies even without
polarimetry.  This means that we can unambiguously determine spin
frequencies in the class as a whole, since nearly all well-studied IPs
show both spin and beat cycles in the emission lines [e.g.\ Hellier
(1997a) and references therein, an exception being TV~Col, where
neither cycle is convincingly detected in the optical (Hellier 1993a;
Augusteijn \etal\ 1994)].

Secure identification of the spin and beat frequencies strengthens the
classification of IPs using the idea that disc-fed accretors show
X-ray pulses (predominantly) at the spin cycle while discless
accretors show X-ray pulses (predominantly) at the beat cycle. Based
on this, we would assign \voph\ as the sole secure member of the
discless category, and a dozen others to the disc-fed category
(e.g.\ Hellier 1991; 1998). An unclear case is TX~Col, in which the
spin pulse dominates at some epochs and the beat pulse at others
(Norton \etal\ 1997; Wheatley 1999).

\voph\ has a 3.42-h orbit and is the highest-field IP at 
9--27 MG. It is likely that this system is on the edge of disc 
formation since we can account for our observations only if there is a 
substantial ring-like structure of blobs circling the white dwarf. 
If the field were a little lower, or the orbit slightly wider (so that 
the angular momentum of the blobs caused them to orbit further out), 
the ring of blobs might turn into an accretion disc. Note that PQ~Gem, 
the only IP with a comparable field estimate (9--21 MG; V\"ath \etal\ 1996; 
Potter \etal\ 1997) has a wider orbit of 5.19 h (Hellier 1997a). It 
appears to accrete though a disc, judging by the dominance of the 
X-ray spin pulse (Mason \etal\ 1992). It is thus likely that, with
these two systems, we are delineating the conditions for disc formation
in an IP.

The theoretical requirements for disc formation have long
been debated (e.g.\ Hameury \etal\ 1986; Lamb \&\ Melia 1988). 
The issue centers on the
three distances \rmin\ (the radius of closest approach to the white
dwarf of a ballistic stream from the Lagrangian point), the larger
circularisation radius, \rcirc\ (where orbiting material has the same
angular momentum that it had at the Lagrangian point), and \rinner\ (the radius
at which the disc is magnetically disrupted). [We distinguish \rinner\
from the \rmag\ used above since the radius is different for
spherical, stream, and disc flows; see Warner (1995) for a fuller
discussion and formulae for these radii.]

It is easy to form a disc with \rinner\,$<$\,\rmin, since the stream
could freely orbit outside the magnetosphere, circularise at \rcirc,
and form the disc. It is not possible to form a disc with
\rinner\,$>$\,\rcirc\ since the stream has insufficient
angular momentum to orbit at these radii. The intermediate case,
\rcirc\,$>\,$\rinner\,$>$\,\rmin, is less clear, but is the case
applying to most IPs. For example, if we assume that PQ~Gem has a
red-dwarf mass of 0.5 \msun\ (appropriate for its 5.19-h orbital
period), a white-dwarf mass of 0.7 \msun, and that its magnetosphere
corotates with the Keplerian orbit at \rinner, then, given the 833-s
spin period, we have \rinner\,\appro 1.1\pten{10}\,cm, which is 
larger than \rmin\ (\appro 0.6\pten{10}\,cm) and close
to \rcirc\ (\appro 1.1\pten{10}\,cm). 

\voph\ could be showing us the mechanism for disc formation under these
circumstances. In this discless system a ring of diamagnetic blobs
orbits inside the magnetosphere. In doing so, the blobs cross field lines
and lose angular momentum to the field, and so eventually accrete. 
However, if the field were weaker, as in most IPs, the drag on the 
diamagnetic blobs would be lower, and their inward spiral would be slower.
The blobs might accumulate, screening the field from each other, and so 
accumulate further in a runaway process leading to the formation of a disc.

\section{Conclusions}
$\bullet$ \voph\ is at a low inclination, estimated as 10\deg. We only ever
see the upper magnetic pole. 

$\bullet$ There is no accretion disc in this system. The stream-fed accretion 
flips from pole to pole on the beat cycle, and this can be seen 
directly in the line profiles. However, only \appro 25 per cent
of the accreting material participates in this motion.

$\bullet$ The rest of the material appears to be circling the white dwarf,
perhaps in the form of diamagnetic blobs.  Such blobs may have been a 
precursor to disc formation in other intermediate polars.

$\bullet$ The spin-cycle variations of the emission lines are accounted
for by a simple model of stream-fed accretion. The best match between
model and data is found for ditances 5--10 \rwd\ from the white dwarf.
The offset between the
magnetic and spin axes of the white dwarf must be at least 30\deg.

%\section*{Acknowledgments}

\end{document}